\documentclass[pra,amsmath,twocolumn,showpacs]{revtex4}

\renewcommand{\figurename}{\textbf{Figure}}
\usepackage{amssymb}
\usepackage{graphicx,xcolor}
\usepackage{epigraph}
\usepackage{csquotes}
\usepackage{amsmath,amssymb,stmaryrd}
\usepackage{amsmath,graphicx}
\begin{document}
\def\la{{\langle}}
\def\u{\hat U}
\def\U{\hat U}
\def\B{\hat B}
\def\C{\hat C}
\def\D{\hat D}
\def\Q{\hat Q}
\def\c{\uppercase}
\def\l{\ell}
\def\fb{\overline F}
\def\wb{\overline W}
\def\nl{\newline}
\def\R{\text{Re}}
\def\H{\hat{\mathcal H}}
\def\lm{\lambda}
\def\lmu{\underline\lambda}
\def\q{\quad}
\def\t{\tau}
\def\n{\\ \nonumber}
\def\ra{{\rangle}}
\def\Ep{{\mathcal{E}}}
\def\omga{{\epsilon}}
\def\t{{\tau}}
\def\up{\uparrow}
\def\dn{\downarrow}
\def\h{\hat{H}}
\def\BB{{\textbf{\textit B}}}
\def\AA{{\textbf{\textit A}}}
\def\BBB{{\bf{ B}}}
\def\AAA{{\bf{A}}}
\def\aa{{\bf{ a}}}
\def\bb{{\bf{b}}}
\title{A minimalist's view of quantum mechanics}
%\shorttitle{} %Insert here a short version of the title if it exceeds 70 characters

\author{D. Sokolovski$^t{1,2}$}
% \and S. Author\inst{1} \and T. Author\inst{2}}
%\shortauthor{F. Author \etal}

\affiliation{$^1$ Departmento de Qu\'imica-F\'isica, Universidad del Pa\' is Vasco, UPV/EHU, Leioa, Spain}
\affiliation{$^2$ IKERBASQUE, Basque Foundation for Science, E-48011 Bilbao, Spain}
\begin{abstract}
We analyse a proposition which considers quantum theory as a mere tool for calculating probabilities 
for sequences of outcomes of observations made by an Observer, 
who him/herself remains outside the scope of the theory. 
Predictions are possible, provided a sequence includes at least two such observations.
Complex valued probability amplitudes, each defined for an entire sequence of outcomes, are attributed to Observer's reasoning,
 %rather than seen as objective properties of physical world, 
and the problem of wave function's collapse 
is dismissed as a purely semantic one. Our examples include quantum "weak values", and a simplified version 
of the "delayed quantum eraser". 
\end{abstract}
%\pacs{03.65.Ta}{Foundations of quantum mechanics}
%\pacs{03.65.AA}{Quantum systems with finite Hilbert space}
%\pacs{03.65.UD}{Entanglement and quantum nonlocality}
\maketitle
%
% repeat the \author\address pair as needed
%
%\author {D. Sokolovski$^{a,b}$} 
%\email {dgsokol15@gmail.com}
%\author {S.A. Gurvitz$^c$}
%\author {J. Siewert$^{a,b}$}
%\author {L.M. Baskin$^c$}
%\author {J. G. Muga$^{a,d}$}
%\affiliation{$^a$ }
%\affiliation{$^b$ }
%\affiliation{$^c$ 
%Department of Particle Physics and Astrophysics, Weizmann Institute of Science, Rehovot, 76100, Israel}
%\begin{abstract}
%\noindent
%\textbf{We .}

 \date\today
%\end{abstract}

%
% insert suggested PACS numbers in braces on next line
%
%\pacs{PACS number(s): 03.65.Ta, 73.40.Gk}
%\pacs{03.65.-w, 03.65.Yz, 03.75.Nt}%\maketitle
%\begin{center}
%\today
%\end{center}
%
%\large{
%
%\section{Introduction}
\epigraph{...there must be a
certain conformity between nature and our thought.}{H. Hertz}

Unlike classical mechanics, which has its conceptual issues largely settled by the end of the 19-th century \cite{Hertz},
quantum theory appears to need an interpretation, which would go beyond mere statement of its mathematical
apparatus. 
One of the reasons for this is the peculiar use of complex valued wave functions or, more generally, amplitudes,
needed whenever one wishes to evaluate frequencies (probabilities) with which the observed events would occur 
under identical circumstances.
Much depends upon how these amplitudes are considered mere computational tools, or essentials 
attributes of the physical world..
\newline
Present suggestions range from the pragmatic Copenhagen interpretation (see \cite{Cop} and Refs. therein)  to the highly subjective QBism (see \cite{QB}),
and include the Bohmian mechanics (see \cite{Bohm}), Everett's many worlds theory (see \cite{MW1}), and the consistent histories approach (see  \cite{CHA}), to name but a few.
Among the issues at stake is the role and the place of a conscious Observer, famously brought into the discussion as \enquote{Wigner's friend} \cite{Wig}. Another such issue is the  \enquote{collapse} of the wave function (see \cite{Coll} and Refs. therein),
i.e., a sudden change in the observed system's state, apparently not described by the Schroedinger equation. 
\newline
While many of the mentioned interpretations, \cite{Cop} -\cite{CHA}, each in its own way, aim at a global description 
of physical world, our objective is somewhat more modest.
\newline
The purpose of this paper is to try to formulate a basic approach, 
consistent  with the elementary quantum mechanics, 
as well as with particular assumptions about the general issues mentioned above. 
We would like the approach to be as simple as possible, and the assumptions 
to be few. A certain amount of philosophising is, therefore, unavoidable.
%\newline
%and could be tested, corrected, or discarded later.
%o be tested on a larger scale later.  
We start by asking what could one expect from 
quantum theory? A possible answer can be found in Feynman's Lectures \cite{FeynL}, and we reproduce it here in full:
{\it \enquote*{ So at the present time we must limit ourselves to computing probabilities. We say at the "present time", 
but we suspect very strongly that it is something that will be with us forever - that it is impossible to beat this  puzzle - that it is the way nature really}} is.' 
\newline
In the above quote \enquote{we} clearly refers to conscious Observers.  The probabilities, on the other hand, tend to be mentioned in literature in at lest three different contexts. Objective probabilities are related to frequencies, with which events occur \cite{Obj}, 
subjective probabilities describe degree of one's belief \cite{Finetti}, and abstract probabilities, satisfying
Kolmogorov's axioms \cite{Kolm},  are a mathematical concept.  
In what follows, we will be interested mostly in the first option, thus assuming that the purpose of the theory is to predict relative frequencies of events, or of series of events, should the same experiment(s) be repeated a large number of times under the same conditions.
\newline
What kind of events should then be considered? 
Physics is an empirical science, so the \enquote{events} must refer to objective experiences, i.e., accessible in principle to any number, or to all, conscious Observers. (Such is, for example, observation of the moon in the night sky, whereas one's dream about the moon must fall outside the remit of physical sciences.) 
\newline
At this point one faces a further choice to be made. 
Either quantum theory is so universal as to describe the nature of human 
consciousness, and of life in general,
or it is a tool, specifically tailored 
%somas to correctly predict 
to and constrained by the limitations of the Observer's perception.
The Observer is either a subject of the theory, or its user,
%In the former case, the Observer is a subject, placed within the scope of the theory. In the latter case he/she/it 
%must be a user, a client,
 in which his/her place is outside the theory's scope.
\newline
For an imperfect analogy, consider a community of mobile phone users whose ability, for reasons unknown, is limited to enacting applications on the set's screen. After some trying, the users will be able to compile a rule book, similar to a basic operation manual. But, unable to look inside the set, they
will ultimately arrive at the level beyond which no further understanding of telephony's principles is possible. 
%and any guess, based on other everyday experiences, is likely to be erroneous. 
Conversely, although 
these basic rules will say something about how the users communicate, they will provide little insight 
into the origin of human consciousness. The analogy is imperfect since, unlike the physical world, a smartphone was made by user's peers, and more detailed descriptions of the set's design, Maxwell's theory, and  the network's  infrastructure are, in principle, available.  
\newline
%explain the the evolution and organisation of a human body. 
Thus, a choice needs to be made, and in  the following, we will opt for the second proposition.
An assumption that the make up of the world can be known in its entirety is a strong, and a relatively recent one.
Arguably, the fact that a theory inevitably arrives at the level where no further explanation is possible, 
and the nature simply {\it is} as it is, may point towards the existence of phenomena, inaccessible to the Observer's
experience (cf. the awkward analogy of the previous paragraph). 
%What cannot be perceived in principle, cannot be a subject of physical sciences, and 
There is little doubt that human observers have only limited perceptive powers, e.g., an ability of directly observing 
(leave other four senses aside) only surfaces of objects in three dimensional coordinate space.
(Hence the need to equip a measuring device with a pointer,
%For this reason,  it is often assumed, as we will do in what follows,  that a measuring device should have a pointer,
 whose spatial 
displacement encodes the value of the measured variable.)
\newline
Another reason for excluding the Observer from the remit of the theory is that no one has so far observed  a state of human consciousness, while little is said to be known even about the consciousness of ants or trees \cite{FeynL}. (This is not to be confused with observation of physical or biochemical 
precesses in a live organism, accessible to direct or indirect measurements  \cite{vN}). Even if the required observational technique could, at some stage, be found, the state of one's consciousness will be accessible to all except the conscious person, caught in a bad progression of being aware of being aware ... of being aware of his/hers own state.
This, in turn, contradicts the earlier requirement that physics should deal only in phenomena, 
accessible to all in equal measure. The old view that {\it \enquote{inner life of an individual is ... extra-observational  by its very nature}} \cite{vN},
and that quantum mechanics should not try to describe the Observer entirely, 
is currently regaining its popularity. One example can be found in a recent paper by Frauchieger and Renner \cite{OBS1}, although 
valid critique of the analysis was later given in \cite{OBS2}.
%(To a reader,  wishing to accuse us of pompous philosophising, we are ready to apologise 
%for the shortcomings of the presentation.  One needs, however, to hold an opinion about the general issues 
%outlined above, and we give no apologies for bringing them up.)
%With the Observer set apart in his/hers/its client role, and the theory reduced to predicting causal connections and correlations between his/hers/its experience, we need to concentrate on the experiences themselves. 

Having adopted a view that quantum theory is {\it for}  rather than {\it about} conscious Observers, 
we can move on to more practical issues.
We will do so by analysing the case where the events, perceived by an Observer, are the results of observations, made on a elementary quantum system, with which the theory 
associates a Hamiltonian operator $\h$, and a Hilbert space of a finite dimension $N$. An Observer may want to measure 
a variable $\mathcal {C}$, represented by a hermitian operator $\C$, with eigenstates 
$|c_n\ra$, and eigenvalues $C_i$, some of which can be degenerate.
% It is usually accepted that a quantum system is described, at a all times,  by its state (wave function) \cite{Dirac}. 
Quantum theory postulates that an accurate measurement of $\mathcal {C}$ must yield one of the discrete values $C_i$. The  outcome of a measurement 
cannot be predicted with certainty, but the probability of obtaining a $C_i$ is given by
\begin{eqnarray} \label{11}
P(C_i)=\la \psi(t)| \hat{\pi}(C_i)|\psi(t)\ra,
\end{eqnarray}
where $|\psi(t)\ra$ is the state in which the system is at the time of measurement,
$ \hat{\pi}(C_i)=\sum_{n=1}^N |c_n\ra  \Delta (C_i-\la c_n|\hat C| c_n\ra)\la c_n|$ is the projector onto the state, or a subspace, corresponding to the value $C_i$. [Above we have introduced $\Delta(X-Y)$, which equals $1$ if $X=Y$, and $0$ otherwise.]  
\newline
A closer look at  eq. (\ref{11}), which descries a {\it single} measurement of $\mathcal {C}$, shows that, in fact, 
it establishes a correlation between {\it two} Observer's experiences. 
An Observer must first determine that the system is indeed in $|\psi(t)\ra$, prior to the measurement, 
and only then evaluate the odds on having the outcome $C_i$. 
The first step can be made by preparing the system with the help of an apparatus, controlled by the Observer,  or by measuring, at some $t_0 < t$,
another variable,  $\mathcal {B}$ with non-degenerate eigenvalues $B_i$, so that obtaining a $B_j$ 
also helps establish that  $|\psi(t_0)\ra=|b_j\ra$. Either way, with $|\psi(t)\ra=\u(t,t_0)|\psi(t_0)\ra$,  where 
$\u(t,t_0)=\exp\left (-i\int_{t_0}^t \h (t')dt'\right )$, 
%is the system's evolution operator, 
eq. (\ref{11}) now yields a conditional probability for obtaining first 
$B_j$ and later $C_i$, 
\begin{eqnarray} \label{21}
P(C_i)=P(C_i\gets B_j)\equiv
% \q\q\q\q\q\q\n
\sum_{n=1}^N \Delta ( C_i-C_n)|A(c_n \gets b_j)|^2.
 %\la \psi(t)| \hat{\pi}(C_i)|\psi(t)\ra.
\end{eqnarray}
A complex valued quantity
\begin{eqnarray} \label{31}
A(c_n \gets b_j)\equiv \la c_i|\u(t,t_0)|b_j\ra,
\end{eqnarray} 
is a Feynman's transition amplitude \cite{FeynH} for a system which starts in $|b_j\ra$ at $t_0$ and ends up in $|c_n\ra$ at $t$.
\newline
Importantly, the sequence $C_i\gets B_j$ cannot be reduced further, e.g., to predicting the statistics of measuring $B_i$ on its own.
There are two compelling reasons why the concept of the state of a system, previously not a subject to an Observer's experience, 
can have little physical meaning. Suppose, Alice receives a spin-$1/2$ from a completely unknown source.
One way to determine the state it is in would be produce a large number of its identical copies, and perform measurements 
on the ensemble, created in this manner. But this is forbidden by the no-cloning theorem \cite{NOCLO}, since the task cannot be performed
by means of a unitary evolution, the only kind of evolution allowed by quantum theory. 
Alternatively, Alice could make a single measurement, but the result will depend on the choice of the measured operator, 
and cannot, therefore, reveal the true state of the system before Alice's intervention. 
(See also  \cite{Merm}, for a proof that a measured value cannot pre-exist its measurement, and must be {\it produced} in the course of it.) 
%These considerations 
We are, therefore, encouraged to
shift the focus of our attention from the wave function $|\psi(t)\ra$ in eq. (\ref{11}), to the transition amplitude $A(c_n \gets b_j)$ in eq. (\ref{31}), related the correlations between several events, experienced by the Observer.
% So far this is in line with our earlier attempts to place Observers 
%outside the domain of applicability of quantum mechanics.
\newline
%We note also that
However,
 the two-measurements case (\ref{21})-(\ref{31})  is not fully representative of the problem at hand, and we will turn to  sequences in which three or more quantities  $\mathcal Q^\l$, $\l=1,2,...,L$, are measured  times 
$t_\l$,  $t_{\l+1}>t_\l$, with the possible outcomes 
 $Q^1_{i_1}, Q^2_{i_2}, ..., Q^L_{i_L}$ (apologies for the cumbersome notations).
To predict  the probability  
 of a given series of outcomes,  $P(Q^L_{i_L}...\gets Q^1_{i_1})$, one must construct
  complex valued probability amplitudes
 for all possible scenarios, add them as appropriate, and take the absolute square of the result  \cite{FeynL}. 
This rule needs to
 take into account the degeneracies
of eigenvalues $Q^\l_{i_\l}$ of the operators $\Q^\l$, representing the quantities $\mathcal{Q}^\l$, 
and can be summarised as follows. 

{\it (I) Virtual (Feynman) paths.} First, one needs to introduce $L$ complete basis sets $\{|q^\l_{n_{\l}}\ra\}$, 
$n_\l=1,2,...,N$, in which the operators $\Q^\l$ are diagonal.  Connecting the states at different times $t_\l$,
yields $N^L$ virtual paths $\{q^L_{n_L}...\gets q^2_{n_2}\gets q^1_{n_1}\}$, each endowed with its own
probability amplitude, 
\begin{eqnarray} \label{51}
A(q^L_{n_L}...\gets q^2_{n_2}\gets q^1_{n_1}) =
\la q^L_{n_L}|\u(t_L,t_{L-1})|q^{L-1}_{n_{L-1}}\ra \times \n
...\la q^3_{n_3}|\u(t_3,t_{2})|q^{2}_{n_{2}}\ra
%\n
\la q^2_{n_2}|\u(t_2,t_{1})|q^{1}_{n_{1}}\ra.
\end{eqnarray}
These paths are the elementary building blocks, from which the observable probabilities 
%$A(Q^L_{i_L}...\gets. Q^1_{i_1}$ in eq. (\ref{41})
will later be constructed. 

{\it (II) Superposition principle.} We will start with the case where first measured eigenvalue, $Q^1_{i_1}$, is non-degenerate, thus allowing for one to one correspondence $Q^1_{i_1} \leftrightarrow |q^{1}_{i_{1}}\ra$,  and return to a more general case in (IV) below.
%in order to define unambiguously the first state of the sequence, $|q_{i_1}\ra$.
%since failure to do so will prevent one from initiating the sequence. 
Other eigenvalues may, or may not, be degenerate, but different rules apply to 
the \enquote{present}, at the last time $t=t_L$, and the \enquote{past} at $t=t_\l$, $1<\l < L$.
 If several eigenstates correspond to a  "past" value 
$\Q^\l_{i_\l}$, one must allow for the interference between the paths, not distinguished by the measurement.
In this case,  the amplitude for obtaining such a value, and ending up in a state $|q^L_{n_L}\ra$, is given by 
\begin{eqnarray} \label{51b}
A(q^L_{n_L}...\gets Q^{\l}_{i_{\l}}...\gets Q^1_{i_1}) =
\sum_{n_2,n_3,...,n_{L-1}=1}^n \prod_{\l=2} ^{L-1}\q\n
\Delta\left (Q^\l_{i_\l}-\la q^\l_{n_\l}|\Q^\l|q^\l_{n_\l}\ra\right )
A(q^L_{n_L}\gets q^{L-1}_{n_{L-1}}...\gets q^1_{i_1}).
%\la q^L_{n_L}|\u(t_L,t_{L-1})|q^{L-1}_{n_{L-1}}\ra....\la q^3_{n_3}|\u(t_3,t_{2})|q^{2}_{n_{2}}\ra
%\la q^2_{n_2}|\u(t_2,t_{1})|q^{1}_{n_{1}}\ra
\end{eqnarray}
However, no interference is allowed for the paths, leading to different 
%(i.e., distinguishable in principle \cite{Feynl})
final states, even if the last observed (\enquote{present}) value $Q^L_{i_L}$ is degenerate \cite{FeynL}.
In this case we have 
\begin{eqnarray} \label{41}\nonumber
P(Q^L_{i_L}\gets Q^{L-1}_{i_{L-1}}....\gets q^1_{i_1})=\n
\sum_{n_L=1}^N \Delta\left (Q^L_{i_L}-\la q^L_{n_L}|\Q^L|q^L_{n_L}\ra\right )\\
|A(q^L_{n_L}\gets Q^{L-1}_{i_{L-1}}...\gets Q^1_{i_1})|^2,
%\q\q\q\q
%|A(Q^L_{i_L}...\gets. Q^1_{i_1})|^2.
\end{eqnarray}
which reduces to a simple Born rule for a non-degenerate $Q^L_{i_L}$, 
\begin{eqnarray} \label{51a}
P(Q^L_{i_L}....\gets Q^1_{i_1})=|A(q^L_{i_L}...\gets Q^1_{i_1})|^2.
\end{eqnarray}
The rule demonstrates, for example, that {\it at present} a particle cannot be at two different locations in space.
Suppose that (we moved from finite-dimensional  systems to point particles in one dimension), at $t=t_L$, one measures a projector onto an interval $[a,b]$, $\hat \pi_{[a,b]}=\int_a^b |x\ra \la x| dx$. There is no amplitude for being inside $[a,b]$, whose absolute square gives the probability to obtain an eigenvalue $1$. Rather, there are probabilities for being at each location inside the interval, whose sum yields the odds on obtaining this eigenvalue.
Not so, if $\hat \pi_{[a,b]}$ is measured {\it in the past}, at some $t_\l <t_L$, where one has to define an amplitude for passing through the entire interval according to eq. (\ref{51b}), and take its absolute square, as prescribed by eq. (\ref{41}). This is true in every representation, determined by the 
observations one wishes to make. 
\newline
We will return to the need for a distinction  between the past and the present in the next paragraph, after
noting that if Alice, wishes to test the theory, she can prepare a statistical ensemble by measuring 
a $\Q^1$, selecting those systems, for which the outcome is a non-degenerate eigenvalue $Q^1_{i_1}$, 
and proceeding to measure the values  of the remaining $\mathcal Q^\l$. The gathered statistics will then be described by eqs. (\ref{41}).

{\it (III) Causality and consistency.} Causality requires that the observations made in future may not 
affects the results already experienced. Indeed, it is easy to check that ignoring the outcomes, obtained 
at $t=t_L$, restores the probabilities (\ref{41}), for a shorter sequence
$\{ Q^{L-1}_{i_{L-1}}....\gets Q^1_{i_1}\}$,
\begin{eqnarray} \label{61}
P(Q^{L-1}_{i_{L-1}}....\gets Q^1_{i_1})=
\sum_{i_L}P(Q^L_{i_L}\gets Q^{L-1}_{i_{L-1}}....\gets Q^1_{i_1}).
%|A(Q^L_{i_L}...\gets. Q^1_{i_1})|^2.
\end{eqnarray}
The rule is also consistent, in the sense that to add one more measurement 
of $\Q^{L+1}$ at $t_{L+1} > t_L$, one should simply relegate the moment $t_L$ to the past, 
and consider Feynman paths 
$\{q^{L+1}_{n_{L+1}}\gets  q^L_{n_L}...\gets q^1_{n_1}\}$
with the amplitudes 
\begin{eqnarray} \label{71}
A(q^{L+1}_{n_{L+1}}\gets  q^L_{n_L}...\gets q^1_{n_1})=\la q^{L+1}_{n_{L+1}}|\u(t_{L+1},t_L)
|q^{L}_{n_{L}}\ra \times \n
A(q^L_{n_L}...\gets q^1_{n_1})\q\q\q
\end{eqnarray}
Equation (\ref{71})  helps to provide some insight into the form of eq.  (\ref{41}). Suppose, at $t_L$ 
one measures an operator $\Q^L$, whose eigenvalues are degenerate. It is then possible, 
without altering the probability of the previous sequence of outcomes, to measure an operator $\Q^{L+1}$, 
diagonal in one of the bases, in which $\Q^L$ is also diagonal. If the eigenvalues of $\Q^{L+1}$ 
are all distinct, and the measurement is made immediately after $t_L$, $t_{L+1}\to t_L$, 
the first factor in the r.h.s. of eq. (\ref{71}) is  a Kronecker delta, $\la q^{L+1}_{n_{L+1}}|\u(t_{L+1},t_L)
|q^{L}_{n_{L}}\ra\to \la q^{L}_{n_{L+1}}|q^{L}_{n_{L}}\ra=\delta_{n_{L+1}n_L}$.
Inserting (\ref{71}) into eq. (\ref{51b}) (with $L$ replaced by $L+1$), applying the Born rule (\ref{51a}), 
and using (\ref{61}), yields eq. (\ref{41}) for the probability of observing a sequence 
$\{ Q^L_{i_L}\gets Q^{L-1}_{i_{L-1}}....\gets Q^1_{i_1}\}$. This illustrates Feynman's assertion \cite{FeynL}
that scenarios, which can be distinguished {\it in principle} (in this case, by a future more detailed measurement), 
are always {\it exclusive}. In particular, there can be no interference between the paths leading to orthogonal final states.
 
 {\it (IV)  Inconclusive preparation and consistency.}
 Suppose next that the first measurement yields an $1< m\le N$-degenerate  value $Q^1_{i_1}$, 
 with which one associates an $M$-dimensional sub-space of the system's Hilbert space, spanned 
 by a basis set $|u_m(Q^1_{i_1})\ra$, $m=1,2,...,M$. This information is not sufficient for assigning to the system a particular initial state, and  Alice, who still wishes to create a statistical ensemble
 (or asked to guess the next outcome given her incomplete knowledge \cite{Finetti}),
 must make an additional assumption about what that state should be. Consistent with the result $Q^1_{i_1}$, the system, prepared  in any of the states $|u_m(Q^1_{i_1})\ra$.
Assuming that $m$-th choice is made with a probability $\omega_m\ge 0$, $\sum_{m=1}^M\omega_m=1$, Alice obtains
 %which gives
 \begin{eqnarray} \label{41x}
P(Q^L_{i_L}...\gets Q^{\l}_{i_{\l}}....\gets Q^1_{i_1})=\n
\sum_{m=1}^M \omega_m P(Q^L_{i_L}...\gets Q^{\l}_{i_{\l}}....\gets u_m)
%\q\q\q\q
%|A(Q^L_{i_L}...\gets. Q^1_{i_1})|^2.
\end{eqnarray}
where $P(Q^L_{i_L}\gets Q^{L-1}_{i_{L-1}}....\gets u_m)$ is the probability (\ref{41}) for the system, which was prepared in 
a state $|u_m\ra$. With all possible choices of $\omega_m$, and of orthonormal bases spanning the $M$-dimensional subspace, Alice has many options. One does, however, stand out.  With no other information available, she can decide to give all $|u_m(Q^1_{i_1})\ra$ equal weights, thus choosing 
 \begin{eqnarray} \label{41y}
\omega_m=1/M.
\end{eqnarray}
Now the probabilities (\ref{41x}) no longer depend on a particular choice of the basis $|u_m(Q^1_{i_1})\ra$, 
since $\sum_{m=1}^M |u_m\ra\la u_m|=\sum_{m=1}^M |u'_m\ra\la u'_m|=\hat \pi(Q^1_{i_1})$.
(Note that Alice could as well consider {\it all} states in the subspace to be equally probable. A demonstration is straightforward 
for $M=2$, where the states  can be parametrised by the polar and azimuthal angles, and the integration of the corresponding 
projectors over the entire Bloch sphere yields one half of the unity operator.) 
\newline
Furthermore, with the choice (\ref{41y}) made, the rule is consistent in the sense that if  $\mathcal Q^1$ is a constant quantity, 
$\Q^1 = \lambda \hat I$, and the first measurement yields no information whatsoever, 
 % and the first result is certain, 
  $P(Q^L_{i_L}...\gets Q^{\l}_{i_{\l}}....\gets u_m)$
reduces to the probability of a shorter sequence,
 \begin{eqnarray} \label{41z}
P(Q^L_{i_L}...\gets Q^{\l}_{i_{\l}}....\gets Q^1_{i_1}=\lambda)=\n
%\n
 P(Q^L_{i_L}...\gets Q^{\l}_{i_{\l}}....\gets Q^2_{i_2}),\q\q
%\q\q\q\q
%|A(Q^L_{i_L}...\gets. Q^1_{i_1})|^2.
\end{eqnarray}
as if the first measurement, whose outcome is certain, were not made at all.

{\it (V) Composites and separability.} With the help of the above, one can treat observations, made on a system of interest (labelled $S$, with a Hamiltonian $\h_S$),
seen as a part of a larger composite system+environment (labelled $E$, with a Hamiltonian $\h_E$), whatever this environment  might be. 
The full Hamiltonian is now given by $\H = \h_S+\h_E+\h_{int}$, where the last term describes the interaction between 
the $S$ and $E$. If, for example,  the environment is a system in a $K$-dimensional Hilbert space, the eigenvalues 
$Q_i(S)$ of an operator $\Q(S)$, representing a system's variable $\mathcal Q(S)$, are at least $K$-fold degenerate. 
The probability of a series of outcomes of observations made on the $S$  are, therefore, given by eqs. (\ref{51})-(\ref{41}), 
with $\u(t_{\l+1}, t_\l)=\exp\left (-i\int_{t_\l}^{t_{\l+1}} \H(t')dt'\right )$. It is easy to check that if the system is completely isolated from the environment, so that $\h_{int}=0$ and $\u(t_{\l+1}, t_\l)=\exp\left (-i\int_{t_\l}^{t_{\l+1}} \h_S(t')dt'\right )\otimes
\exp\left (-i\int_{t_\l}^{t_{\l+1}} \h_E(t')dt'\right )$, after summing over the degeneracies one recovers eqs. (\ref{51})-(\ref{41}) for the system only, with 
$\u(t_{\l+1}, t_\l)=
%\u^S(t_{\l+1}, t_\l)\equiv  
\exp\left (-i\int_{t_\l}^{t_{\l+1}} \h_S(t')dt'\right )$.
Note that here we have also assumed that at $t=t_1$ the result of the first measurement corresponds to a composite's product state $|q^1_{i_1}(S)\ra\otimes |q^1_{j_1}(E)\ra$. A measurement of a more general collective quantity, 
$\Q(S+E)$, may yield $Q^1_{i_1}(S+E)$, which would leave
 the composite 
 %at $t_1$ 
 in an entangled state  
%$|q^1_{i_1}(S+E)\ra=\sum_{j_1=1}^N\sum_{k_1=1}^K\beta_{j_1k_1}|q(S)^1_{j_1}\ra |q(E)^1_{k_1}\ra$. 
$|q^1_{i_1}(S+E)\ra=\sum_{j_1=1}^N\beta_{i_1j_1}|q^1_{j_1}(S)\ra\otimes|\phi_{j_1}(E)\ra$,
$\la \phi(E)_j|\phi(E)_j\ra=1$. 
If no further interaction between $S$ and $E$ is possible, 
%for the probability 
% of a sequence of  outcomes of measurements made only on the $S$ 
application of eqs. (\ref{51})-(\ref{41}) to each term of the sum yields
 \begin{eqnarray} \label{81}
P(Q^L_{i_L}(S)... \gets \Q^\l_{i_\l}(S)
%\gets Q^{L-1}_{i_{L-1}}(S)
....\gets Q^1_{i_1}(S+E))=\q\q\n
\sum_{n_L=1}^N \Delta \left (Q^L_{i_L}(S)-
\la q^L_{n_L}(S)| \Q^L(S)|q^L_{n_L}(S)\ra\right )\n
%\times\q\q\q \n
 \times \sum_{j,j'=1}^N \beta_{i_1j'}^*\beta_{i_1j}
 \la \phi(E)_{j'}|\phi(E)_j\ra\q\q\q\q\q\q\n
 % \times \n
\times A^*(q^L_{n_L}(S)... \gets \Q^\l_{i_\l}(S)....\gets q^1_{j'}(S))\q\q\q
\n
\q\q\times A(q^L_{n_L}(S)... \gets \Q^\l_{i_\l}(S)....\gets q^1_{j}(S))\q\q\q	
%\sum_{n_L=1}^N \Delta\left (Q^L_{i_L}-\la q^L_{n_L}|\Q^L|q^L_{n_L}\ra\right )
%|A(q^L_{n_L}\gets Q^{L-1}_{i_{L-1}}...\gets Q^1_{i_1})|^2,
%|A(Q^L_{i_L}...\gets. Q^1_{i_1})|^2.
\end{eqnarray}
 which simplifies to 
 \begin{eqnarray} \label{81a}
 P(Q^L_{i_L}(S)....\gets Q^1_{i_1}(S+E))
 =\n
 \sum_{j=1}^N|\beta_{i_1j}|^2 P(Q^L_{i_L}(S)....\gets q^1_j(S) )
 \end{eqnarray}
 in a special case where $\phi_j(E)\ra$ are 
 orthogonal, $\la \phi(E)_{j'}|\phi(E)_j\ra=\delta_{jj'}$, and the system (S) can be said to start in a state
 $|q^1_{j}(S)\ra$ with a probability $|\beta_{i_1j}|^2$. 

 Equations (\ref{51})-(\ref{41}) and (\ref{41x}) can be re-written in a compact and, perhaps, more familiar form
 \begin{eqnarray} \label{81c}
P(Q^L_{i_L}... \gets \Q^\l_{i_\l}
%\gets Q^{L-1}_{i_{L-1}}(S)
....\gets Q^1_{i_1})= \q\q\q\n
%\q\q\q\q\n
%\la q^1_{i_1}| \hat \pi(Q^1_{i_1},t_1)\hat\pi(Q^2_{i_2},t_2) \times\n
% ...\hat\pi(Q^{L}_{i_L},t_L)....\times\hat\pi(Q^2_{i_2},t_2)\hat\pi(Q^1_{i_1},t_1)|q^1_{i_1}\ra\q
 Tr \{ 
 %q^1_{i_1}| \hat \pi(Q^1_{i_1},t_1)
 \hat\pi(Q^2_{i_2},t_2
 ) 
 %\times\n
 ...\hat\pi(Q^{L}_{i_L},t_L)....\times\hat\pi(Q^2_{i_2},t_2)
% \hat\pi(Q^1_{i_1},t_1)
 \hat \rho (Q^1_{i_1})\},
\end{eqnarray}
where, the Heisenberg representation, $\hat \pi (Q^\l_{i_\l}, t_\l)\equiv \u^{-1}(t_\l,i_1)\hat \pi(Q^\l_{i_\l})\u(t_\l,i_1)$ is the projector
onto the eigen-subspace, associated with an outcome $Q^\l_{i_{\l}}$ 
%of the operator $\Q^\l=\sum_{i_\l}
%Q^\l_{i_\l}\hat \pi(Q^\l_{i_\l})$,
and $\rho (Q^1_{i_1})=
 \sum_{m=1}^M |u_m\ra\omega_m\la u_m|$ 
 is the system's density operator\cite{vN}.
 % after the first measurement., 
 %if the first  (preparatory) measurement yields an $M$-degenerate eigenvalue of $\Q^1$..
 Similar strings of projectors appear, for example, in the consistent histories approach (CHA) \cite{CHA}, but there are important differences. Firstly, while the CHA aims to be a general theory, which includes observers, 
 we put an Observer outside the theory's scope. Secondly, for us eq. (\ref{81c}) is a derived result, and the primary and most basic quantities are the probability amplitudes (\ref{51}) and (\ref{51b}).
 \newline
As an example where this difference is important, 
consider the case where the system is \enquote{pre- and post-selected} in the states $|q^1_{i_1}\ra$ and $|q^3_{i_3}\ra$ at $t_1$ and $t_3$,
and  in eq. (\ref{81}) the role of environment is played by a von Neumann pointer \cite{vN}, set up to measure some $\Q^2$ at
$t_1<t_2<t_3$. The measurement can more accurate, or less accurate, yet any information about the system, gained from the pointer's final position, will have to be expressed in terms of an amplitude $A(q^3_{i_3}\gets Q^2_{i_2}\gets q^1_{i_1})$ in eq. (\ref{51b})\cite{DSMAR}.
 If $\Q^2=|q^2_m\ra\la q^2_m|$ is a projector onto a state $|q^2_m\ra$, and the coupling to the pointer is small (the accuracy of the measurement is poor), the average shift of the pointer, $f$, turns out to be given by 
 \begin{eqnarray} \label{91}
\la f\ra \approx \R \left [\frac{A(q^3_{i_3}\gets q^2_{m}\gets q^1_{i_1})}{\sum_{n=1}^NA(q^3_{i_3}\gets q^2_{n}\gets q^1_{i_1})}\right ] =\n
\R\left [ \frac{\la q^3_{i_3}(t_2)|\Q^2|q^3_{i_1}(t_2)\ra}{\la q^3_{i_3}(t_2)|q^1_{i_1}(t_2)\ra}\right ],\q\q
\end{eqnarray}
where $|q^3_{i_3}(t_2)\ra\equiv\u^{-1}(t_3,t_2)|q^3_{i_3}\ra$ and $|q^1_{i_1}(t_2)\ra\equiv\u(t_3,t_2)|q^1_{i_1}\ra$.
The last expression in eq. (\ref{91}) was first obtained in \cite{Ah1}, 
where  the complex valued fraction in brackets was called \enquote{the weak value (WV) of the operator $\Q^2$}.
Written in this way, a WV looks like a physical variable of a new kind \cite{WVrev}, whose physical significance 
is still discussed in the literature (see, for example \cite{Matz1}, \cite{Matz2}).
However the first expression in the r.h.s. of eq. (\ref{91}) identifies it as a previously known renormalised Feynman amplitude (or a weighted sum of such amplitudes if a more general $\Q^2$ is inaccurately measured) \cite{DS1} -\cite{DS6}.
The problem is, little known about 
the probability amplitudes, apart from their relation to the observable frequencies, discussed above. Until, or,  unless a deeper insight into 
the physical meaning of quantum amplitudes is gained, such an  inaccurate \enquote{weak} measurement will remain merely
an exercise in recovering the values of transition amplitudes from a response of a system to a small perturbation \cite{DSMAR}.
\newline
Our first simple example  concerns a pair of entangled spins, travelling in opposite directions, 
and can be found in the Appendix A. 
 As a further illustration of the use of the transition amplitudes, we revisit, in its simplest version, the 
 "delayed choice quantum eraser experiment \cite{Scully}. Figure 1a sketches 
 %a situation in which 
% initial measurement of an operator $\hat B$ prepares
a primitive double-slit experiment, in which  
 a two-level system (S) (a spin-1/2), which an observed outcome  $B_1$ has left  in a state  $|b_1\ra$, is subjected to a later measurement of an operator  $\hat C(S) = C_1|c_1\ra\la c_1|+C_2|c_2\ra\la c_2|$ at some $t=t_2$. The final state 
 $|c_1\ra$, playing the role of a point on the screen, can be reached via passing, at a $t_1<t_2$, through a pair of orthogonal states
 $|\up\ra$ and  $|\dn\ra$, representing the two slits. The two virtual paths in the two-dimensional Hilbert space (Fig.1a, solid)  interfere, and the probability 
 to have $C_1(S)$ is given by (spin has no own dynamics)
\begin{eqnarray} \label{zz1}
P(C_1\gets B_1)= |A_1+A_2|^2,\q\q\q\n
A_{1}=\la c_1|\up\ra\la \up|b_1\ra, \q A_2=\la c_1|\dn\ra\la \dn|b_1\ra. 
\end{eqnarray}
In Fig.2b, initial measurement of a collective variable $\hat B(S+E)$ entangles the system with a two-level "environment" (E), whose orthogonal states are $|+\ra$ and $|-\ra$. As before, $\hat C(S)$ is measured at $t_2$, and then an environment's variable $\hat D(E) = D_1|d_1\ra\la d_1|+D_2|d_2\ra\la d_2|$ is measured at a $t_3>t_2$.
Four relevant virtual paths in the now four-dimensional Hilbert space (solid lines in Fig.1b) are endowed with
probability amplitudes
\begin{eqnarray} \label{zz2}
A_{I}=\la d_1|+\ra A_1, \q A_{II}=\la d_2|+\ra A_1,\n
A_{III}=\la d_1|-\ra A_2, \q A_{IV}=\la d_2|- \ra A_2, 
\end{eqnarray}
Using the rules (I)-(VI), for the probabilities of the sequences of outcomes shown in Fig.1c, we easily find
\begin{eqnarray} \label{zz3}
P(D_1\gets C_1\gets B_1)=|A_I+A_{II})|^2\q\n
P(D_2\gets C_1\gets B_1)=|A_{III}+A_{IV}|^2,
\end{eqnarray}
and 
\begin{eqnarray} \label{zz4}
P'(C_1\gets B_1) \equiv P(D_1\gets C_1\gets B_1)
+\n
P(D_2\gets C_1\gets B_1)
= |A_1|^2+|A_2|^2\q
%\q\q\q\q\q\q
\end{eqnarray}
Much of the interest in the above scheme stems from the fact that while there is no interference term 
$\sim A^*_1A_2$
in Eq.(\ref{zz4}), this term reappears in $P(D_1\gets C_1\gets B_1)=|A_1+A_2|^2/2$, if we choose
$|c_1\ra =[|+\ra +|-\ra]/\sqrt 2$. It is tempting to conclude that the coherence between the paths $1$ and $2$, 
that was lost after measuring $\hat C(S)$ at $t_2$ is somehow restored if the second system, (E), is found in $|d_1\ra$.
All the more surprising is that this seems to happen after the outcome $C_1$ has already been observed. 
However, Fig.1b shows that here we are not comparing like with like. In Fig.1b, the interference term
is controlled by the magnitudes of the amplitudes of two virtual paths, $I$ and $II$, which connect  states in a different four-dimensional Hilbert space of the composite system, and the argument cannot be reduced to to a discussion of the individual 
paths shown in Fig.1a.  Quantum theory does its job of calculating probabilities for the outcomes in Fig.1c in an
explicitly causal manner,  and,  we suspect, cannot be asked to do more than that. 
(For other recent attempts at "demystifying" the delayed eraser experiment we refer the reader to Refs.
\cite{DM1}, \cite{DM2}.)
%%%%%%%%%%%%%%%%%%%%%
\begin{figure}
\includegraphics[angle=0,width=8.8 cm, height= 5.5cm]{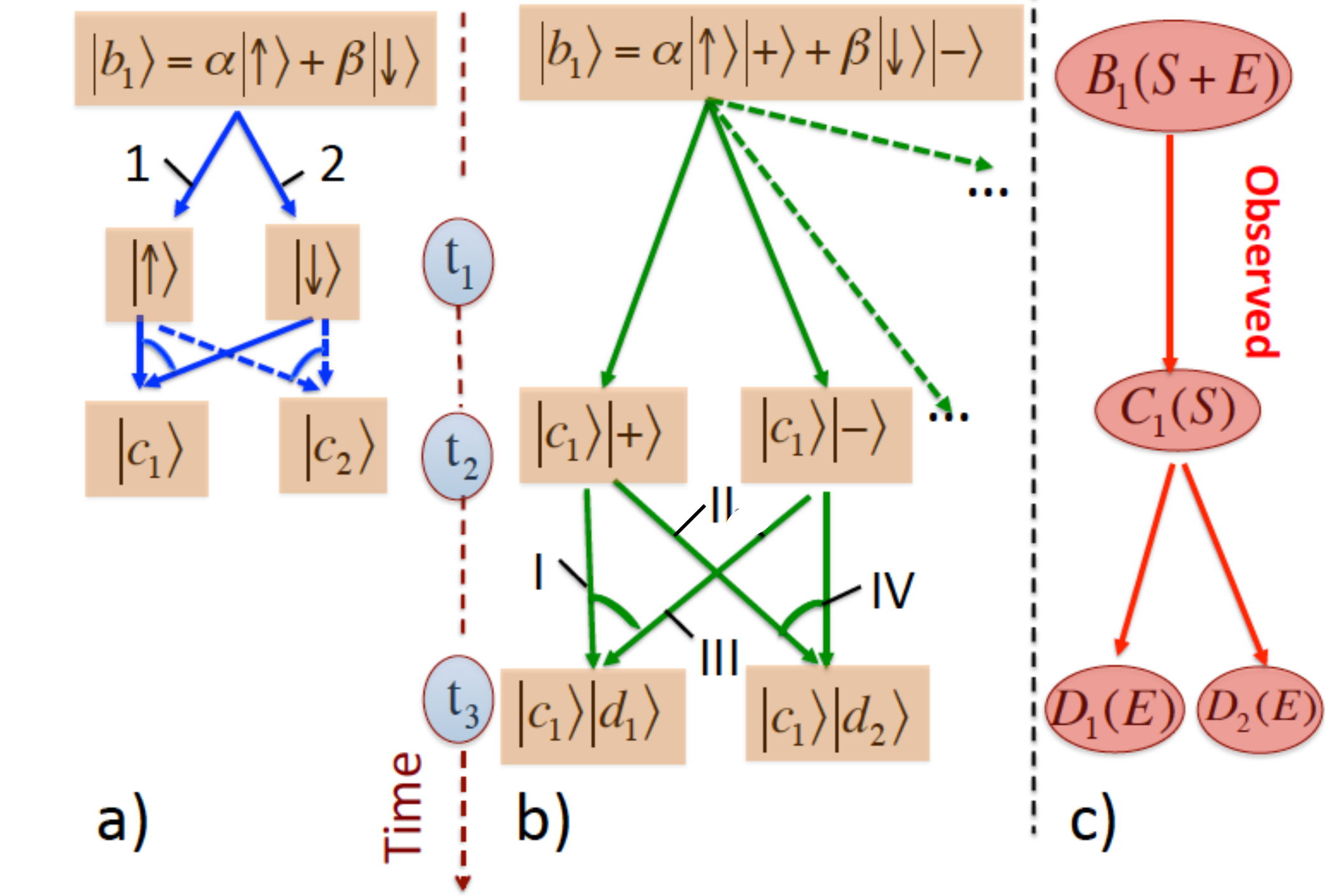}
\caption{a) A primitive "double-slit problem", with four virtual paths connecting states in a two-dimensional Hilbert space of a spin (S). No observation is made at $t_1$, and the paths joined by an arc continue to interfere. 
b) A different "double-slit problem", with virtual paths  (only 4 of the 16  are shown) connecting states in a four-dimensional  Hilbert space of the composite $(S+E)$. A measurement made on the spin 
at $t_2$ does not destroy interference between paths leading to the same final outcomes.  
c) The observed outcomes: what is measured at $t_3$ cannot affect the net probability of obtaining 
$C_1$ at $t_2 < t_1$.}
%(Only four of the 16 virtual paths are shown.}   
\label{fig.1}
\end{figure}
%\newline

We can now sum up the \enquote{minimalist view}, advertised in the title. 
%The proposition is as follows.
Quantum theory is a tool, allowing a conscious Observer to predict statistical correlations between
the results of  two or more of his/hers observations, first of which is needed to \enquote{prepare the system}.
With a particular series of results in mind, 
%(more than 1 outcome are needed) 
he/she may reason about its likelihood by
associating with each  outcome a state, or states, in a Hilbert space, including all systems which interact with  each other during the time interval considered. A probability amplitude for the {\it entire} series is then constructed using the prescriptions (I)-(V), and taking its absolute square yields the required probability. 
Such attributes of the theory as amplitudes, Hilbert spaces, 
operators, and Hamiltonians are essentially Hertz's  \enquote{symbols of external objects, formed by ourselves}, \enquote{whose consequences are always the necessary consequences in the nature of the thing pictured} \cite{Hertz}. (A brief  discussion, which we expect to be consistent with H. Hertz's original view is given in the Appendix B.) Observer's main effort then goes into identifying a system's Hamiltonian $\h$, and the operators $\Q^1$, $\Q^2$,... $\Q^L$, whose  spectra contain all possible observational outcomes .
As a tool, tailored to Observer's limited abilities, the theory is unable to progress beyond certain explanatory level, 
where it must admit, as in the opening quote, that nature simply {\it is} this way.
For the same reason, Observer's conscience is not a valid subject of quantum theory, which must loose its pretence  (if any)
at explaining the world in its entirety, and give way to other complementary endeavours. 
\newline
Several other remarks can be in order. 
Firstly, the symbolic status of probability amplitudes does not preclude that their values can, under certain conditions, be deduced from 
the measured probabilities \cite{DSMAR}. Indeed, this was done, for example, in the experiments reported in \cite{EXP1} and \cite{EXP2}. 
\newline
Secondly, different sets of measurements, e.g., of the quantities $\mathcal Q^\l$ and $\mathcal Q'^\l$, made on the same system, may produce essentially different statistical ensembles \cite{Feyn1},\cite{DS5}.
For example, less detailed probabilities (some of $\Q^l$'s eigenvalues are degenerate) cannot be 
obtained by adding the most detailed ones (obtained for a $\Q'^l$, whose eigenvalues are all non-degenerate).
This is, of course, a more elaborate version of a double slit experiment, where the price of knowing
the way a particle has taken is the loss of the interference pattern on the screen.
\newline Finally, so far we made no mention of the collapse of the wave function. 
The possibility of avoiding this issue altogether, is precisely the point we intend to make here.
An Observer, whose reasoning only requires him/her to evaluate certain matrix elements in an abstract space, may discard the "collapse problem" as a purely semantic one.
It is, of course, possible to argue that in  eq. (\ref{11}) the evolution of the state $|q^{L-1}_{n_{L-1}}\ra$ is mysteriously  interrupted 
at $t=t_L$, but it is equally possible not to enter into this argument at all.
%this remark would add nothing new to the viewpoint described above, and could as well be omitted. 
Conceptual economy from not having to worry about the fate of the wave function can be significant, 
as one avoids dealing with a universe which splits
every time a measurement is made as it happens, 
for example, in Everett's many worlds (MW) picture \cite{MW1}. Curiously, in 1995 M. Price  \cite{MW} polled physicists
in order  to determine the level of support for the MW approach, and counted Feynman among the supporters.
We note that Feynman's support must have been lukewarm at best. In \cite{Feyn1} one reads:
{\it \enquote{ Somebody mumbled something about a many-world picture, and that many-world picture says that 
the wave function $\psi$ is what's real, and damn the torpedoes if there are so many variables, $N^R$.
All these different worlds and every arrangement of configurations are all there just like our arrangement of 
configurations, we just happen to be sitting in this one. It's possible, but I'm not very happy about it.}  }
\newline
To conclude, we note that the proposed viewpoint imposes strict limits 
and, if adopted, will have implications for such concepts as the "universal wave function" \cite {UWF}, for attempts to  construct 
a quantum theory where no special role is given to an observer, \cite{Bohm}-\cite{CHA}, or for collapse-related theories of quantum mind \cite{QM}. None of these matters are trivial, and cannot be dismissed out of hand.
The format of this Letter does not allow for detailed comparisons, so our purpose here was to 
articulate a maximally reduced view, which can later be extended, modified, or abandoned.  For instance, it is possible that the simple model, used to illustrate it, will not suffice when dealing with extremely large or complex systems, or where the relativistic effects of various kinds need to be taken into consideration. 
% It is also possible that current lack of agreement about the foundations of quantum mechanics will be overcome, 
 It is also possible that, contrary to the Feynman's
quote at the beginning of this article, quantum analysis has not yet arrived at its explanatory limit.  
If so, a more sophisticated theory will have to provide a further insight into the meaning of the transition probability amplitudes,
which so far have played  the role of basic elements of a quantum analysis. 
%\begin{eqnarray} \label{201}
%P(\Q(S)^3_{i_3}\gets \Q(S)^2_{i_2} \gets q(S+E)^1_{i_1})=|A(q^3_{3}q^2_{1}
%%%%%%%%%%%%%%%%%%%%%%%%%%%%%%%%%%%%%%
\section{Appendix A. A simple example: two spins $1/2$ in an entangled state}
%As a further illustration of the use of the transition amplitudes, 
 %of a description based on the transition amplitudes (\ref{51}), 
 Here we revisit a well known case of two spin-$1/2$ particles, $1$ and $2$, 
prepared in one dimensional wave packet states, $|\phi(\pm p_0,1,2)\ra$,  moving in opposite directions, and polarised up and down the $z$-axis, respectively. The spins do not interact with each other, there are no external fields,
and  the full Hamiltonian acts only on the spatial degrees of freedom, $ \H=\hat p_1^2/2\mu+\hat p_2^2/2\mu$.
Reversing the spins' directions and adding up the two states one can entangle the spins in a state
(for clarity we use the tensor product sign $\otimes$)
\begin{eqnarray} \label{1 01}
|q(S+E)^1_{i_1}\ra =\bigg [|\up_z,1\ra \otimes|\dn_z,2\ra-|\dn_z,1\ra \otimes|\up_z,2\ra\bigg ]\n 
\otimes |\phi(p_0,1\ra\otimes |\phi(-p_0,2)\ra/\sqrt 2,
\end{eqnarray}
where the \enquote{system} (S) consists of two spins, and the \enquote{environment} (E) includes the spatial degrees of freedom.
By a time $t_2$, after two wave packets have moved well away from each other, Alice measures the component the first spin along a 
direction, $\vec n=(\theta, \varphi=0)$, $\Q(S)^2=\hat \sigma_n(1)$, $Q(S)^2_{i_2}=\pm 1$. At a later time $t_3\ge t_2$ Bob 
measures his spin along a different direction,  $\vec n'=(\theta', \varphi'=0)$, $\Q(S)^3=\hat \sigma_n(2)$, $Q(S)^3_{i_3}=\pm 1$, and we are interested in the four probabilities $P(\Q(S)^3_{i_3}\gets \Q(S)^2_{i_2} \gets q(S+E)^1_{i_1})$.
\newline 
Since there is no spin-orbit interaction, no matter how far apart the wave  packets are, we can use (V) to eliminate the environment, 
and consider only degrees of freedom of the two spins, prepared in a state, 
\begin{eqnarray} \label{201}
|q^1_{j_1}(S)\ra =\bigg [|\up_z,1\ra \otimes|\dn_z,2\ra-|\dn_z,1\ra \otimes|\up_z,2\ra\bigg ]/\sqrt{2}.
\end{eqnarray}
There are four basis states, 
\begin{eqnarray} \label{201x}
|I\ra =|\up_n,1\ra\otimes|\up_{n'},2\ra, \q |II\ra =|\up_n,1\ra\otimes|\dn_{n'},2\ra,\q\n
|III\ra =|\dn_n,1\ra\otimes|\up_{n'},2\ra, \q |IV\ra =|\dn_n,1\ra\otimes|\dn_{n'},2\ra,
\end{eqnarray}
and sixteen virtual paths $\{K \gets J\gets q^1_{j_1}\}$, $J,K=I,II,III,IV$ shown in Fig.2.
However, only four of them have non-zero amplitudes, 
\begin{eqnarray} \label{201y}
A(K\gets J\gets q^1_{j_1})=\la K| J\ra\la J|q^1_{j_1})\ra =  \delta_{JK}\la J|q^1_{j_1}\ra,
\end{eqnarray}
These four paths lead to orthogonal filial states, and cannot interfere.
A simple calculation yields
the probabilities for Alice's and Bob's outcomes
\begin{eqnarray} \label{201a}
P(1\gets 1 \gets q(S+E)^1_{i_1})=\q\q\q\q\q\q\q\q\q\n 
P(-1\gets \-1 \gets q(S+E)^1_{i_1}))=
\sin^2\left (\frac{\theta-\theta'}{2}\right )/2\q\n
P(1\gets -1 \gets q(S+E)^1_{i_1})=\q\q\q\q\q\q\q\q\n 
P(-1\gets \ 1 \gets q(S+E)^1_{i_1}))=
cos^2\left (\frac{\theta-\theta'}{2}\right )/2.
\end{eqnarray}
%represent exclusive, rather than interfering, alternatives.
If Alice and Bob choose to measure along the same axis, $\theta=\theta'$, they are guaranteed to find their respective spins pointing in the opposite directions, $P(1\gets -1 \gets q(S+E)^1_{i_1})=P(-1\gets \ 1 \gets q(S+E)^1_{i_1}))=1/2$.
(This is true even if the distance between the wave packets is so large, that light cannot travel it 
in $t_3-t_2$ seconds, and  the information about Alice's result cannot reach Bob in time.)
%[The scheme, however, is well known to prevent super-luminal transfer of information.
%If Alice encodes her message into the choice of the direction $\theta$, the already mentioned no-cloning theorem 
%would prevent Bob from creating an ensemble of identical spins, and in this way finding out what that direction was.
%If Alice shares with Bob an ensemble on entangled states, according to (\ref{201}) Bob will find half of his spins up any direction $\theta'$, for any choice of $\theta$ Alice might make.]
Needless to say, we haven't provided a new "explanation" for the entanglement phenomenon, since 
the probabilities in eqs. (\ref{201a}) can be obtained directly from (\ref{81c}).
We did, however, provide a more detailed illustration of the fact that in elementary quantum mechanics  correlations between parts of a system cannot depend on an environment, with which the system does not interact. There is simply no provision for such a dependence, 
even when the environment represents spatial positions of the system's parts.  
\begin{figure}
\includegraphics[angle=0,width=8.5cm, height= 5cm]{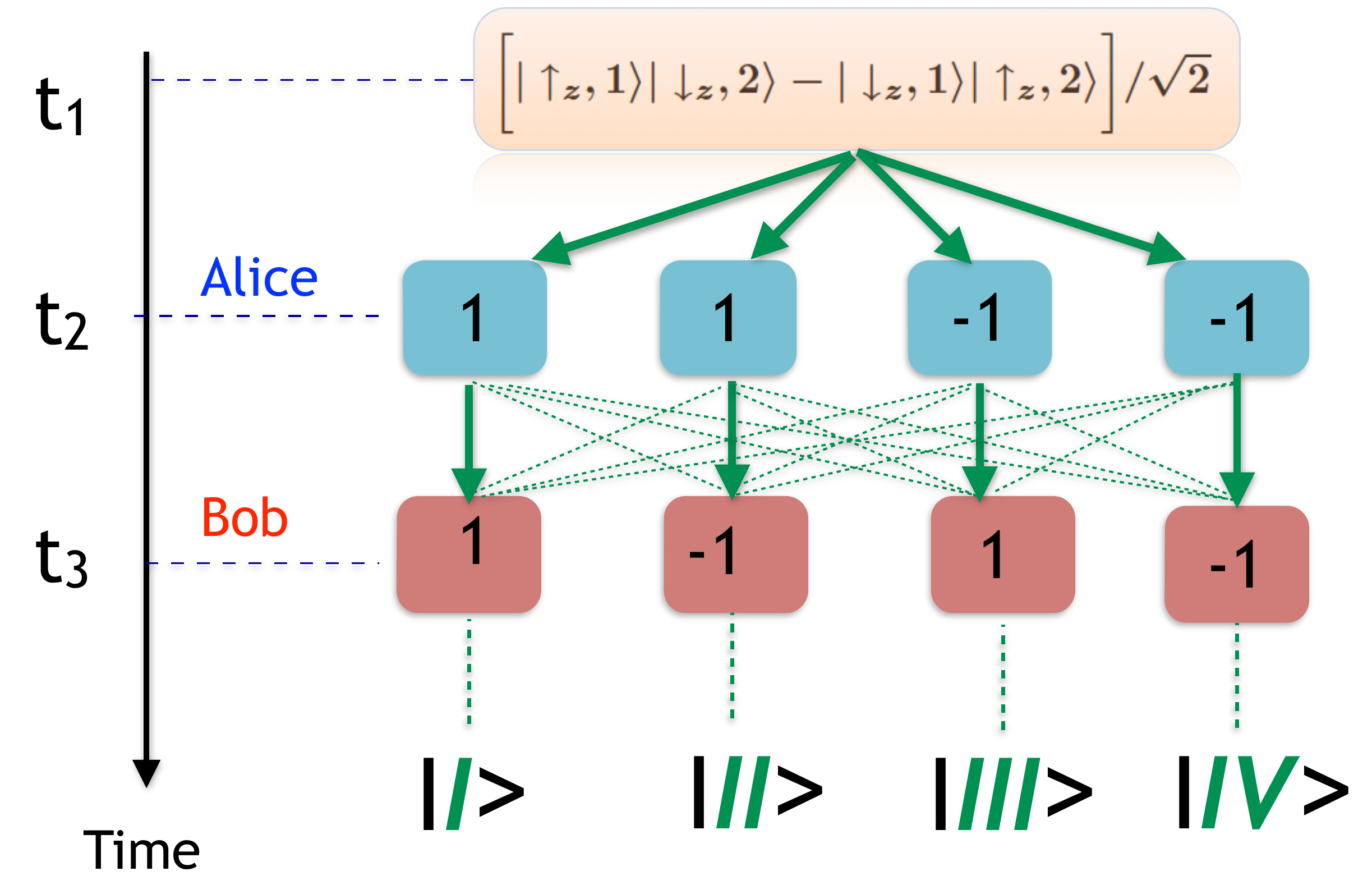}
\caption{Two spins-1/2 are prepared in an entangled state (\ref{201}). Alice and Bob measure 
their spins along the directions $\vec{n}$ and $\vec{n}'$, respectively.
Of the sixteen virtual paths, passing through states in Eqs.(\ref{201x})
%$|\up_n,1\ra \otimes|\up_{n'},2\ra$,
%$|\dn_n,1\ra \otimes|\up_{n'},2\ra$,
%$|\up_n,1\ra \otimes|\dn_{n'},2\ra$, and 
%$|\dn_n,1\ra \otimes|\dn_{n'},2\ra$ 
(dots), only four (solid lines) have non-zero amplitudes.
The numbers $\pm 1$ show Alice's and Bob's outcomes.
In the special case $\vec{n}=\vec{n}'$ the amplitudes of the outer paths vanish, 
and Alice's and Bob's spins alway point in opposite directions.}   
\label{fig.1}
\end{figure}

%%%%%%%%%%%%%%%%%%%%%%%%%%%%%%%%%%%%%%%
\section{Appendix B. Observer's abilities and limitations}
In Fig. 3 we tried to sketch a scheme, broadly consistent with the approach of \cite{Hertz}.
In nature a phenomenon $\BB$ always follows (is caused by) a phenomenon $\AA$.
Neither of the two are fully accessible to the Observer ($O$), who, limited by his/her five senses, 
can only perceive some indications of $\AA$ and $\BB$, namely $\AAA$ and $\BBB$.
In order to establish a connection between observations $\AAA$ and $\BBB$, $O$ associates
$\AAA$ with a {\it symbol}  $\aa$, $\AAA\to \aa$ from his/her theoretical toolkit. 
He/she then uses the theory to reason about the consequences of $\aa$, the result being 
a new symbol $\bb$, $\aa \to \bb$. Observer's theory is correct, if $\bb$ corresponds to the observed result 
$\BBB$, $\bb \leftrightarrow \BBB$,  and false otherwise. 
\newline
Observer's toolkit may include mathematical methods and concepts such as space, time, trajectories, forces, 
Lagrangians, Hilbert spaces, wave functions, etc. Next we must answer, at least to ourselves, the following questions.
Do these items reflect the Observer's limited ability to reason about his/her limited experiences 
of an in principle intractable world?
Or does his/her logical reasoning provide a deeper insight into the actual inner working of nature, 
hidden from sensual perception?  Here we follow \cite{Hertz} in assuming the former.
Classical mechanics is an elegant complete theory, yet it is no longer possible to insist that objects
really obey Newton's laws, now that the theory has beed superseded by quantum mechanics.
Quantum theory comes with its own conceptual baggage, and its even more difficult 
to attribute it solely to nature, leaving the human mind outside the picture. 
To use the well known quote by Asher Peres \cite{PERES}:
 {\it \enquote{ Quantum phenomena do not occur in a Hilbert space. They occur in a laboratory}.}

%What we meant is largely explained in the caption of Fig. 14, and here we will try to give a concrete example.
%%%%%%%%%%%%% 
\begin{figure}
\begin{center}
\includegraphics[angle=0,width=7.5 cm, height= 5.5cm]{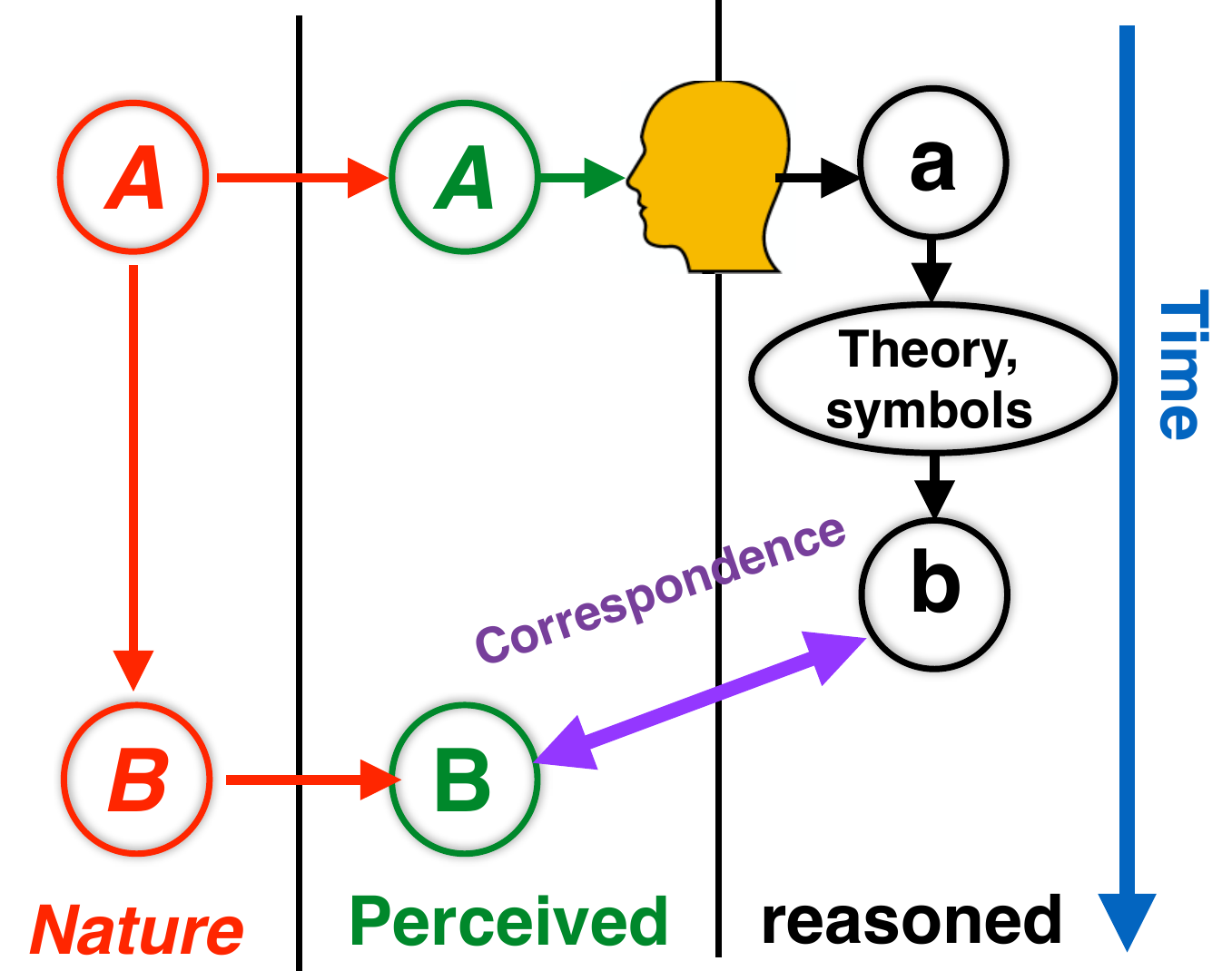}
\caption{In nature,  an event $\AA$ leads to (causes) an event $\BB$.
An Observer ($O$) receives some indication of  $\AA$, $\AAA$ according to his/her perceptive 
abilities. $O$ associates with the observed outcome $\AAA$ a symbol $\aa$  from a 
tool kit of a theory and  evolves into another symbol, $\bb$ by means of logical
operations. If  $\bb$ agrees with the perceived image of the nature's event 
$\BB$, $\BBB$, the theory is correct.
At all times  Observer's knowledge of what really happens in nature remains limited to what he/she
is able to observe. Symbols used by the theory are treated as attributes of Observer's reasoning, 
rather than observer-free attributes of nature.} 
%(Only four of the 16 virtual paths are shown.
 \label{fig.1}
\end{center}
\end{figure}
We conclude with a simple example.
The laboratory setup consist of a copper coil, connected to an battery, a small object, $S$, (a spin-$1/2$ with 
associated magnetic moment), invisible to the naked eye, and meter (e.g., some sort of Stern-Gertach machine), $M$, which after a brief interaction with the spin 
gives one of two results, $\pm 1$. Alice switches the meter on, and something ($\AA$) happens between $S$ and $M$.
Alice can see ($\AAA$) is that the meter's reading is $+1$.  She begins to reason by associating condition of the spin  ($\aa$), with a vector, $|\uparrow_z\ra$ in a two-dimensional Hilbert space. The coil is represented 
by a magnetic field $\mathcal{B}$, directed along the $x$-axis. Alice writes down an evolution operator 
$\u(t)=\cos(\omega t)+i\sin(\omega t)\sigma_x$
%Hamiltonian $\h=\omega \sigma_x$, 
where $\omega \sim \mathcal{B}$ is the Larmor frequency, an
 amplitude for being able to attribute to the spin the same state $|\uparrow_z\ra$
\begin{eqnarray}\label{3a}
A(\uparrow_z \gets \uparrow_z ) = \la\uparrow_z|\u(t)||\uparrow_z\ra = cos(\omega t),   
\end{eqnarray} 
and the probability to see the meter reading $+1$, if its is enacted  after a time $t=2\pi/\omega$  ($\BB$),
\begin{eqnarray}\label{3a}
P(1\gets 1 ) = |A(\uparrow_z \gets \uparrow_z )|^2 = cos^2(\omega t)=1.   
\end{eqnarray} 
 Alice's prediction ($\bb$) that she will always see a result $+1$ 
can be compared with the actual meter's reading ($\BBB$)
 turns out to be true - her theory is correct.
\newline 
The situation is a little more complicated if Alice wants top make predictions for an arbitrary time $t$.
In general, she cannot say what the meter will read in each individual case. 
%For those who believe that science
%can only deal in situations where the $B$ follows a given $A$ with certainty, such prediction fall outside
%of the theory's remit. 
Alice can, however, prepare many copies oh the same setup, switch all meters 
at the same time and select $N>>1$ cased where the result is $+1$.
 Now $\AA$ in Fig.3 is a collective event, 
and Alice's outcome $\AAA$ is seeing $N$ positive meters' reading.
At $t$ all meters are enacted again, and the outcome observed by Alice ($\BBB$) is the ratio $N_+/N$, where $N_+$ is the number of positive readings. 
For large $N$ Alice has $N_+/N \to cos^2(\omega t)$, as predicted by her theory. 
Note that neither Alice, nor other human observer, can know what "really" happens in the lab in full detail.
But a causal link in the sense "if first $\AAA$, then later $\BBB$", has been established for certain phenomena, accessible 
to Alice's five senses.

%\end{eqnarray}

\end{document}